\documentclass[12pt]{article}
\usepackage{amsmath,amsfonts,amssymb,color}
\usepackage{authblk}
\usepackage{graphicx}

\title{Heavy neutral lepton search and $\mu \to e \gamma$ constraints in case of type-I seesaw}
\author[1,2]{Stefano Morisi}
\affil[1]{Dipartimento di Fisica ``Ettore Pancini'', Università degli studi di Napoli ``Federico II'', Complesso Univ. Monte S. Angelo, I-80126 Napoli, Italy}
\affil[2]{INFN - Sezione di Napoli, Complesso Univ. Monte S. Angelo, I-80126 Napoli, Italy}

\date{}

\begin{document}
  \maketitle
  \begin{abstract}
Within type-I seesaw mechanism it is possible to have large (order one) light-heavy neutrino mixing even in case of low right-handed neutrino mass scale (of the order of GeV).
This implies large lepton flavor violation. As example we consider the process $\mu \to e \gamma$ that can have a branching  up to $10^{-8}$ within type-I seesaw  (in contrast with the tiny value $10^{-54}$ expected).  
Such an enhancing of lepton flavor violation  can be used to constraint the parameter space of  long lived particle experiments.

  \end{abstract}

Observation of neutrino oscillation  is an evidence that neutrino are massive and that flavor neutrino states do not coincide with the massive one. The unitary lepton mixing matrix $U_\nu $ connecting the two basis introduced by
Pontecorvo-Maki-Nakagawa-Sakata (PMNS) has been observed esperimentally to be very different from the identity. 
Indeed one of the three angles  parametrizing $U$ is close to be maximal $\sin^2\theta_{23}\sim0.5$ and one is large $\sin^2\theta_{12}\sim0.3$
while the third is small but not zero $\sin^2\theta_{13}\sim0.02$. 
Lepton mixing suggests that in the Standard Model can be present Lepton Flavor Violation (LFV) phenomena like $\mu \to e \gamma$. Early computation of this process mediated by  the three active light neutrinos gives
\cite{Petcov:1976ff,Bilenky:1977du} (see \cite{Lindner:2016bgg,Ardu:2022sbt} for a recent overview)
\begin{equation}\label{Br1}
Br(\mu \to e \gamma) \approx \frac{3 \alpha_e}{32 \pi} \left|  U_{\nu_{13}}^*U_{\nu_{2 3}} \frac{\Delta m_{31}^2}{M_W^2} \right|^2\sim 10^{-54}\,,
\end{equation} 
that is very far from actual experimental sensitivity that is $7.5\cdot 10^{-13}$\,\cite{MEGII:2023ltw}. \\

If the  standard model is extended by means of  $n$  right-handed neutrino $N_k$ (where $k=1,..,n$) having Majorana mass given by $n\times n$ mass matrix $M_N$,
the neutrino mass matrix is a $(3+n)\times (3+n)$ matrix 
\begin{equation}\label{Mnu}
M_\nu=\left(
\begin{tabular}{cc}
$0$ & $m_D$  \\
$m_D^T$ & $M_N$  
\end{tabular}
\right)\,,
\end{equation}
where $m_D=Y_D v$ is the $3\times n$ Dirac mass matrix ($v$ is the standard model vev) and $Y_D$ is the corresponding Yukawa coupling and 
we assume $m_D\ll M_N$.
Without loss of generality we can go in the basis where $M_N$ is diagonal.
The neutrino mass matrix $M_\nu$  is diagonalized by a $(3+n)\times (3+n)$  unitary matrix $V$ given in block form by
\begin{equation}\label{V}
V= \left(
\begin{tabular}{cc}
$U_\nu$ & $U_{\nu N}$  \\
$U_{\nu N}^\dagger U_\nu $ & $\mathbb{I}$  
\end{tabular}
\right)+\mathcal{O} (\theta^2)\,,
\end{equation}
where  $U_{\nu N}=m_D\cdot M_N^{-1}$ is a $3\times n$ matrix that mixes light and heavy neutrino. 
It follows that the $3\times 3$ lepton mixing matrix $U_\nu$ is a sub-block of the unitary matrix $V$  and therefore there is violation of unitarity in PMNS 
that is typically parametrized  by $\theta^2\equiv U_{\nu N} \,U_{\nu N}^\dagger$, see for instance \cite{Antusch:2017pkq}.
Block diagonalizing $M_\nu$ one obtain the well know 
(type-I) seesaw relation for the three light active neutrinos
\begin{equation}\label{ssw}
m_\nu=-m_D\frac{1}{M_N}m_D^T\,.
\end{equation} 
Using this expression naively, namely assuming only one active neutrino with mass $m_\nu$ and one right-handed neutrino with mass $m_N$, it follows that 
\begin{equation}\label{theta2}
 \theta^2 \sim m_\nu / m_N\,,
\end{equation}
that is suppressed even for light $m_N$, indeed $\theta^2\sim [10^{-10}-10^{-25}]$ for $m_N\sim [10^{-1}-10^{14}] $\,GeV.
This estimation does not really change in the $3+n$ realistic case.\\

In the case of type-I seesaw the branching ratio of the process $\mu \to e \gamma$ is given by
\begin{equation}\label{Br2}
Br(\mu \to e \gamma) \approx \frac{3 \alpha_e}{32 \pi} \left|  \sum_{k=4}^{n} (U^*_{\nu N})_{1k} (U_{\nu N})_{2k}  F(x_k) \right|^2 \,,
\end{equation} 
where $x_k=m^2_{N_k}/m_W^2$ and $F(x)=(10-43x+78x^2-49x^3+18x^3\log x+4x^4)/(6(1-x)^4)$ \cite{He:2002pva}.
Because of eq.\,(\ref{Br1}), the contribution from light active neutrino is negligible, so here we consider only heavy right-handed neutrino in the sum of (\ref{Br2}). Therefore in case of 
type-I seesaw naively it is expected that $Br(\mu \to e \gamma) \sim \theta^2$ is  suppressed. 

Even if this suppression  is true in some limit, this is not the most general result,
in fact  $\theta^2$ can be (theoretically) up to $10^{-1}-10^{-2}$ (as long as  $m_D\ll M_N$ is guaranteed). For large $\theta^2$ the branching (\ref{Br2}) is enhanced and can be up to $10^{-8}$ \cite{Casas:2001sr} (see  also \cite{Antusch:2006vwa} for an effective approach). The aim of the present paper is to update  the main idea of \cite{Casas:2001sr} in a different language that can be useful for experiments searching for long lived particles like heavy neutral leptons, see
for instance ANUBIS \cite{Bauer:2019vqk,Shah:2024fpl}, MATHUSLA \cite{Curtin:2018mvb}, SHADOWS \cite{Baldini:2021hfw}, NA62 \cite{NA62:2017rwk,Drewes:2018gkc}, FASER \cite{FASER:2018eoc}, CODEX-b \cite{Gligorov:2017nwh}. 

Heavy neutral leptons (here right-handed neutrino $N$s) can be produced from  $D,\, B$ meson decay, gauge boson $W,\,Z$, standard model Higgs $H$ and top quark. Indeed 
in the minimal type-I scenario with $n$ right handed neutrino, $N_k$ enter in the charged and neutral current that leads to a coupling of $N_k$ with $Z$ and $W$ bosons,
\begin{equation}
\mathcal{L}\supset -\frac{g}{\sqrt{2}} Z_\mu\, \overline{\nu_L}_{\alpha k}\gamma^\mu N_k\, (U_{\nu N})_{\alpha k} - \frac{g}{\sqrt{2}} Z_\mu\, \overline{\ell}_{\alpha k}\gamma^\mu N_k\, (U_{\nu N})_{\alpha k}\,.
\end{equation}
Such a couplings are at the origin of both $N$ production and decay. 
 Then heavy neutral leptons decay quite far from the production point depending on the $U_{\nu N}\sim\theta$ mixing. Being such a mixing quite small in case of heavy neutral leptons, the lifetime can be up to $\tau_N < 0.1s$ 
 (this upper limit come from Big-Bang Nucleosynthesis constraints). As a consequence the decay length can be much bigger then $100\,m$ and so any detector can catch a small fraction of long-lived particle decay.
 For this reason all this experiment try to maximize the distance from the interaction point and the detector. Just to give an idea the distance is about $20\,m$ for ANUBIS and CODEX-b, $200\,m$ for  MATHUSLA
 and  $480\,m$ for FASER. 
In  \cite{Kling:2018wct} has been shown that  the dominant 
branching of heavy neutral lepton $N$ is into hadrons, but decays  into leptons are also possible. 

The rate for  production and decay of $N$ are both proportional to  $U_{\nu N}$. 
The mixing parameters that are typically considered in long lived experiments are 
\begin{equation}\label{U2}
U_\alpha^2 =\sum_{i=1}^3 | (U_{\nu N})_{\alpha \, i} |^2\,, \qquad U^2 =\sum_{\alpha} | U_\alpha |^2 \,,
\end{equation}
where $\alpha=e,\, \mu,\,\tau$.
The sensitivity of heavy neutral lepton experiments 
is typically reported in the $(U_\alpha^2 - m_N)$  or $(U^2 - m_N)$ plane.

To understand the origin of the enhancing of $\theta^2$ we need to go deeply into the detail of type-I seesaw mechanism.
The Dirac Yukawa coupling $Y_D$ can be parametrized in terms of the physical observable, namely the masses of the light active neutrino and the parameters of the PMNS mixing matrix and the right handed masses by means of the Casas-Ibarra parametrization \cite{Casas:2001sr}
\begin{equation}\label{ci}
Y_D=v^{-1} U_{PMNS} \, \sqrt{m_\nu^{diag}}\, R\, \sqrt{M_N^{diag}}\,,
\end{equation} 
where $R$ is an arbitrary complex $3\times n$ orthogonal matrix.
From relation (\ref{ssw}) it is possible to fit  the two  square mass differences if  $n \ge 2$. The minimal case with $n=2$ 
predicts one massless light active neutrino. 
In the following for simplicity we will consider the case $n=2$ and degenerate heavy right-handed neutrino 
$$m_N\equiv {(M_N)}_{11}={(M_N)}_{22}.$$ 
When $n=2$, the matrix $R$ is given by (for normal neutrino mass ordering considered here)
\begin{equation}\label{R}
R=\left(
\begin{tabular}{cc}
0&0\\
$\cos\beta$ & $\sin\beta$\\
$-\sin\beta$ & $\cos\beta$
\end{tabular}
\right)\,,
\end{equation}
where $\beta=x+i \,y$ is an arbitrary complex number.  The value of $\theta^2$  strongly depends on the parameter $y$ while only mildly on the parameter $x$ that for simplicity we assume to be $x=0$. 
The parameter $y$ can be in principle very large as soon as the seesaw regime is preserved, namely $m_D\ll M_N$. In the present analysis we take $0<y<30$. The fact that $\theta^2$ is not suppressed by the neutrino mass  
$m_\nu$ as in eq.\,(\ref{theta2}) is possible only for large values of $y$. If $y$ is large enough the magnitude of the neutrino Yukawa couplings could be of order one even for  $m_N\sim$\,GeV. This seems to be 
in contradiction with common sense\,(\ref{theta2}) but  is a possibility. 
Using large value of $y$ is therefore possible to obtain an enhancing of $Br(\mu \to e \gamma)$. Barring large $y$ is possible in case of low-scale seesaw mechanism, for a review see 
\cite{Boucenna:2014zba}. A study of large lepton flavor violation coming from unitarity violation in case of low-scale seesaw is given for instance in \cite{Forero:2011pc}.

\begin{figure}
\begin{center}
\includegraphics[scale=0.55]{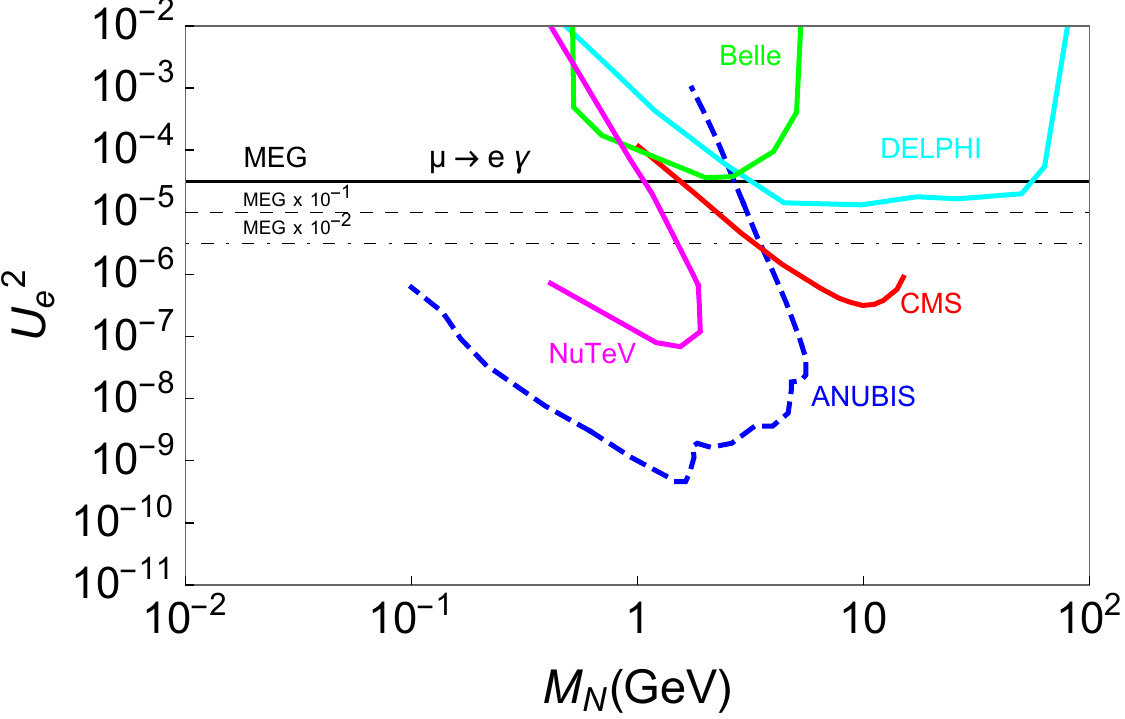}\caption{Light-heavy neutrino mixing parameter $U_e^2$ as a function of the right-handed neutrino mass $m_N$, the horizontal line represents MEG constraint, for details see the text.}\label{figmeg}
\end{center}
\end{figure}

Here we assume  for simplicity as benchmark case the following choice of the parameters appearing in (\ref{ci}): $\sin^2\theta_{23}\sim0.5$,  $\sin^2\theta_{12}\sim0.3$, $\sin^2\theta_{13}\sim0.02$, $m_{\nu 1}=0,\, m_{\nu 2}=\sqrt{\Delta m^2_{12}},\,  m_{\nu 3}=
\sqrt{\Delta m^2_{13}}$ where $\Delta m^2_{12}\simeq 7\cdot 10^{-5}\,eV$, $\Delta m^2_{13}\simeq 2\cdot 10^{-3}\,eV$. Moreover the Dirac and Majorana phases as well as the parameter $x$ are taken to be zero. 
With all these assumptions it follows that 
$Br(\mu \to e \gamma) \sim \theta^2$  depends only by the two free parameters $y$ and $m_N$.
For each set of $y$ and $m_N$ value chosen, the neutrino mass matrix $M_\nu$ is fixed and we obtain numerically the $\mu \to e \gamma $ branching from (\ref{Br2}), $m_D$ from (\ref{ci}) and 
the mixing matrix $U_{\nu N}$ and therefore also the corresponding parameter $U_\alpha^2$ 
and $U^2$ from (\ref{U2}). 

We graph $U_e^2$ as function of $m_N$ marginalizing with respect to $Br(\mu \to e \gamma) < 7.5\cdot 10^{-13}$ (similar graphs can be obtained for $U_\mu^2$, $U_\tau^2$, $U^2$). 
The result is shown in figure (\ref{figmeg}) where
we report with continues lines existing experimental limits (see for instance\,\cite{Drewes:2018gkc}). 
In figure we report with the dashed line in the graph the expected sensitivity of ANUBIS\,\cite{Hirsch:2020klk} taken as representative of long lived particle experiments.  
The horizontal continuous line is the limit coming from MEG. In order to better understand the role of $Br(\mu \to e \gamma)$ in the $(U_e^2,\, m_N)$ plane,
we show with dashed horizontal lines the constraints coming assuming a sensitivity of MEG improved by a factor 10 and 100. \\

The main result of this analysis is that constraints coming from $\mu \to e \gamma$ lepton flavor violation process is in agreement with the actual constraints coming from other experiments. In particular such a limit are of the same order for masses 
$1\,GeV\lesssim m_N \lesssim 80\,GeV$. However above $80\,GeV$ MEG provide new limits. In principle MEG limits can be extended up to grand unified scale, but above 100 GeV the future heavy neutral leptons experiments  are not sensitive. If MEG sensitivity will be improved by a factor 100, then   $\mu \to e \gamma$  constraints could dominate for $m_N\gtrsim 10\,GeV$.

In summary in this analysis we provide a prove of the potentiality of lepton flavor violation in discriminating standard type-I seesaw with the interplay of long lived particle experiments.

%Assuming the relation (\ref{ssw}) it is possible to show that in order to explain  the two  square mass differences observed in neutrino oscillation experiments, it is required that $n_R\ge 2$. The minimal case with $n_R=2$ 
%predicts one massless light active neutrino. Here we focus for simplicity to the case $n_R=2$. 

\end{document}